\documentclass[reprint,amsmath,amssymb,aps,nofootinbib]{revtex4-1}

\usepackage{amsmath,amssymb,graphicx}
\usepackage{hyperref}
\usepackage{float}
\usepackage{verbatim}
\usepackage{subfigure}
\usepackage{ragged2e}
\usepackage{newfloat,algcompatible}
\usepackage{etoolbox}
\usepackage{dcolumn}
\usepackage{bm}
\usepackage{amsfonts}
\usepackage{tabularx}

\input{Qcircuit}

\begin{document}

\title{Resource comparison of two surface code implementations of small angle $Z$ rotations}

\author{Prashant Mishra$^1$, Austin Fowler$^{2,3}$}
\affiliation{
 $^1$Indian Institute of Technology Bombay, Powai, Mumbai, Maharashtra 400 076, India\\
 $^2$Department of Physics, University of California, Santa Barbara, California 93106, USA\\
 $^3$Centre for Quantum Computation and Communication Technology, School of Physics, The University of Melbourne, Victoria 3010, Australia
}


\begin{abstract}
Fault-tolerant $Z$ rotations by $\pi/2^k$ are important as they arise in numerous quantum algorithms, most notably those involving quantum Fourier transforms. We describe surface code implementations of two recently described methods of efficiently constructing these rotations. One method uses state distillation to get low-error $(|0\rangle + e^{i\pi/2^k}|1\rangle)/\sqrt{2}$ states, with each distillation level requiring $2^{k+2}-1$ input states to produce a single purer output state, and uses these distilled states to directly implement $\pi/2^k$ angle $Z$ rotations. The other method is indirect, using sequences of single-qubit Clifford and $T$ gates. We compute and compare the overhead of our surface code implementations of these two techniques. We find that the approximating sequence overhead is less than or equal to direct distillation for $k > 3$ and logical error rates $\lesssim 10^{-12}$.
\end{abstract}

\maketitle

Physically implementing a quantum computer has been an area of intense research over the last decade, and superconducting qubits operating at the threshold  fidelity of the surface code now exist \cite{Fowl12f,Bare13}. Most fault tolerant quantum error correction schemes \cite{Shor95,Cald95,Stea96,Lafl96,Baco06,Knil04c,Fuji10,Brav98,Denn02,Bomb06,Raus07,Raus07d,Bomb10,Katz10,Ohze09b,Bomb10b,Fowl11}, including the surface code, permit low overhead implementation of only gates from the Clifford group, for example $X$, $Z$, $H$, $S$, and CNOT. This is insufficient for universal quantum computation \cite{Brav05,Reic05} as at least one non-Clifford gate is required to achieve quantum universality.

Typically, non-Clifford gates are achieved by first distilling so called magic states \cite{Brav05,Reic05}. Magic states have the property that multiple copies can be distilled to fewer higher fidelity copies of the same state. Magic state distillation is typically high overhead and it is important to search for efficient methods. State distillation can be iterated to reach any fidelity we desire.

Here, we concern ourselves with two recent works of note. The first one \cite{Land13} directly distills states of the form $|\psi_k\rangle=(|0\rangle + e^{i\pi/2^k}|1\rangle)/\sqrt{2}$ and uses them to implement $Z_k=R_Z(\theta_k)$ rotations, where $\theta_k=\pi/2^k$ and $R_Z(\theta_k)=e^{-i\theta_k Z/2}$. The second \cite{Kliu12,Kliu12a,Ross14} uses Clifford and $T$ gates to construct sequences which approximate $Z_k$ rotations. We construct surface code implementations and compute the overhead of executing these non-Clifford gates through these two methods. Even more recent methods of constructing logical rotations \cite{Boch13,Ducl14,Paet13c,Boch14} shall be considered in future work.

\section{State Distillation}

State distillation can be used to reduce multiple copies of certain states to fewer higher-fidelity copies \cite{Brav05,Reic05}. Many distillation protocols have been proposed \cite{Meie12,Brav12b,Jone12}, but here we limit ourselves to the generalized Reed-Muller codes proposed by Landahl and Cesare \cite{Land13}. These can be used to distill the family of states $|\psi_k\rangle$ for $k\geq 1$. These states can then be used to implement quantum $Z_k$ rotations\footnote{Note that $Z_0=Z$, $Z_1=S$, $Z_2=T$.}, which are equivalent, up to a global phase, to
\begin{equation}
Z_k = \begin{pmatrix}
1 & 0 \\
0 & e^{i\pi/2^k}
\end{pmatrix}.
\end{equation}

\begin{figure}
\begin{center}
\input{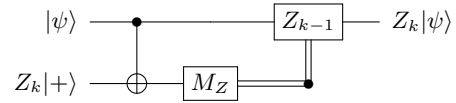}
\end{center}
\caption{Quantum circuit for probabilistically implementing a $Z_k$ gate using $Z_k |+\rangle$ states.}
\label{tpcirc}
\end{figure}

The probabilistic quantum circuit taking $|\psi_k\rangle=Z_k|+\rangle$ as input and implementing $Z_k$ on an arbitrary state $|\psi \rangle$ is shown in Fig.~\ref{tpcirc}. When the measurement reports 0, the desired rotation has been implemented. Half of the time, however, the measurement reports 1, indicating $Z_k^{\dag}$ has been implemented, necessitating an attempt to implement $Z_{k-1}$. This process at worst continues until $Z_1$ is attempted, as $Z_0=Z$ is a Pauli operator and can be accounted for in software.

The $|\psi_k\rangle$ distillation scheme uses a family of generalized quantum Reed-Muller codes $C_k$ and achieves an input-output error relation\footnote{Exact relation is given by the expression \cite{Brav05, Land13} - \[ p_{out}(p) = \frac{1-(1-2p)^{2^{k+1}-1} \left[ 2p(2^{k+2}-1)+(1-2p)^{2^{k+1}} \right]}{2 \left[ 1+(2^{k+2}-1)(1-2p)^{2^{k+1}} \right]} \]}
\begin{equation}
p_{out}(p)= A_kp^3+O(p^4).
\end{equation}

These generalized Reed-Muller codes $C_k$ have the property that a logical $Z_k$ rotation can be implemented using transversal $Z_k^\dag$ rotations (i.e. implemented on each qubit independently). This means we can prepare a surface code Bell pair and then encode one half of the pair as a $C_k$ Reed-Muller code. If we then apply transversal $Z_k^\dag$ to this encoded half, and measure the encoded half transversally in the logical $X$ basis, we can process these measurement results to detect if any errors are present and keep the remaining half only if no errors are detected. The codes $C_k$ are capable of detecting any combination of one or two errors. So the remaining half, if no errors are detected, will have an error probability of order $p^3$.

The precise output error rate depends on the size of the code itself, the complexity of the distillation circuitry, and the strength of error correction used. If we initially assume that sufficiently strong error correction is used to suppress any errors in the distillation circuitry, the number of undetectable combinations of three errors that corrupt the output is shown in Table~\ref{valAK} \cite{Land13}. These numbers can be expressed compactly as
\begin{equation}
A_k = \frac{1-3(2^{k+1})+2^{2k+3}}{3}.
\end{equation}

\begin{table}
\centering 
\setlength{\extrarowheight}{3pt}
\begin{tabular}{|c | c| c| c|} 
\hline\hline 
\hspace{5mm} $k$ \hspace{3mm} & \hspace{0mm} $A_k$ \\ [0.5ex] 
\hline 
\hspace{5mm} 1 \hspace{5mm} &  \hspace{5mm} 7      \hspace{5mm}  \\
\hspace{5mm} 2 \hspace{5mm} &  \hspace{5mm} 35     \hspace{5mm}  \\
\hspace{5mm} 3 \hspace{5mm} &  \hspace{5mm} 155    \hspace{5mm}  \\
\hspace{5mm} 4 \hspace{5mm} &  \hspace{5mm} 651    \hspace{5mm}  \\
\hspace{5mm} 5 \hspace{5mm} &  \hspace{5mm} 2667   \hspace{5mm}  \\
\hspace{5mm} 6 \hspace{5mm} &  \hspace{5mm} 10795  \hspace{5mm}  \\
\hspace{5mm} 7 \hspace{5mm} &  \hspace{5mm} 43435  \hspace{5mm}  \\
\hspace{5mm} 8 \hspace{5mm} &  \hspace{5mm} 174251 \hspace{5mm}  \\
\hline\hline 
\end{tabular}
\caption{Values of $A_k=p_{out} / p^3$ for different values of $k$ (to the most significant order).} 
\label{valAK} 
\end{table}

\begin{figure}
\begin{center}
\input{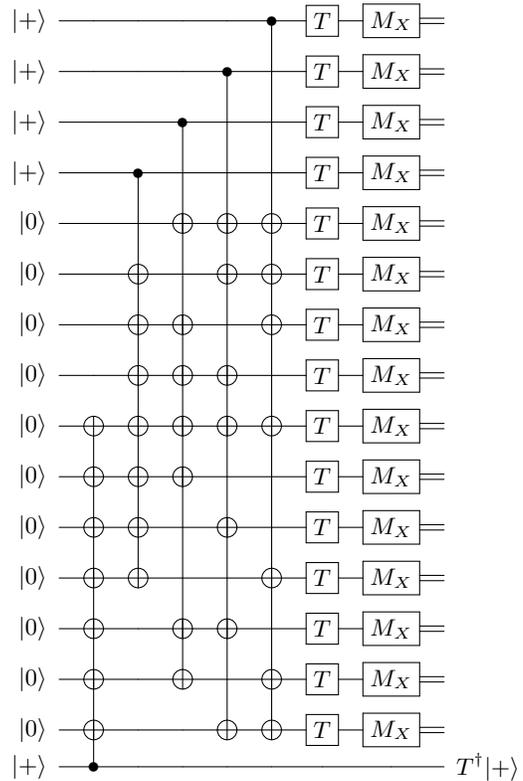}
\end{center}
\caption{Distillation circuit for $T^\dagger |+\rangle$ states. The circuit takes 15 copies of $T|+\rangle$ with error $p$ and gives us one $T^\dagger |+\rangle$ with error approximately $35p^3$. The $T$ gates each consume input states and can be implemented by the recursive circuit shown in Fig.~\ref{tpcirc}.}
\label{z2circ}
\end{figure}

\begin{figure}
\begin{center}
\input{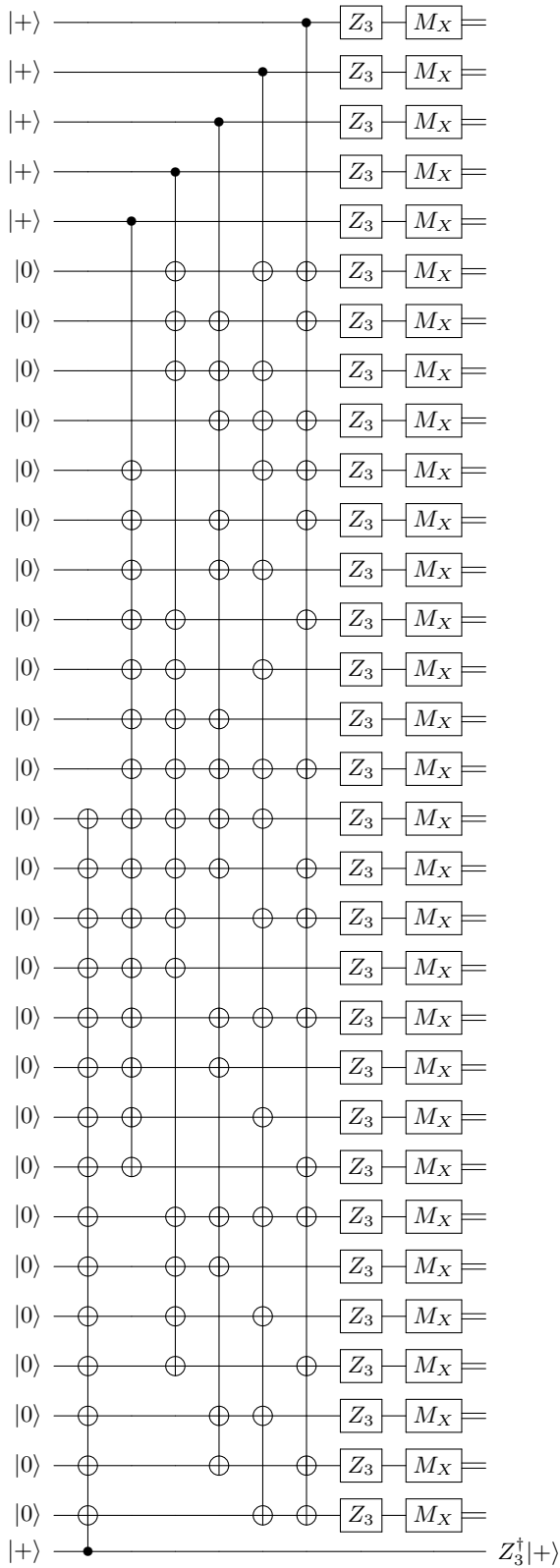}
\end{center}
\caption{Distillation circuit for $Z_3^\dagger |+\rangle$ states. The circuit takes 31 copies of $Z_3 |+\rangle$ with error $p$ and gives us one $Z_3^\dagger |+\rangle$ with error approximately $155p^3$. The $Z_3$ gates each consume input states and can be implemented by the recursive circuit shown in Fig.~\ref{tpcirc}.}
\label{z3circ}
\end{figure}

The quantum circuit used to distill $T^\dagger |+\rangle$ states is shown in Fig.~\ref{z2circ} \cite{Land13}. Each of the $T$ gates is produced using a copy of $T|+\rangle$ of error $p$ and the circuit shown in Fig.~\ref{tpcirc}. As can be seen, Fig.~\ref{z2circ} requires 15 $T|+\rangle$ input states to produce a lower error output state. Similarly, the circuit used to distill $Z_3^\dagger |+\rangle$ states is shown in Fig.~\ref{z3circ} \cite{Land13}. This circuit requires 31 $Z_3|+\rangle$ input states to produce an output of lower error. In general, we need $2^{k+2}-1$ copies of $|\psi_k\rangle$ as input to produce a single better $|\psi_k\rangle$ output. These lower error copies can then be distilled further or used to implement $Z_k$ gates as shown in Fig.~\ref{tpcirc}.

Note that either $|\psi_k\rangle$ or $|\psi_k^\dag\rangle$ can be used to implement $Z_k$ without changing the probability of obtaining the desired gate, which is always 50\%. Note also the recursive nature of Fig.~\ref{tpcirc}, implying the need for lower $k$ states to run $|\psi_k\rangle$ distillation.

We now turn our attention to implementing circuits of the form shown in Fig.~\ref{z2circ} and Fig.~\ref{z3circ} using the surface code. In order to do so, we first provide a brief review.

\section{Topological Quantum Error Correction}

Topological quantum error correction schemes \cite{Brav98,Denn02,Bomb06,Raus07,Raus07d,Bomb10,Katz10,Ohze09b,Bomb10b,Fowl11,Fowl12f} require only nearest neighbor interactions and are desirable because of their high tolerance to error \cite{Raus07,Wang11,Fowl13g}. Topological schemes measure local operators to detect errors. If we consider the directions in which the array of qubits is laid as spatial dimensions and the direction of computational time as a time dimension, a defect is a region in this space-time where we do not perform measurements to detect errors. Computation can be performed by braiding defects. Certain chains of errors connecting defects and rings of errors encircling defects are undetectable and can cause logical errors. Logical errors can be exponentially suppressed by using large-circumference defects well separated from each other or, more specifically, by stretching the defects in all dimensions. The operators measured to detect errors are called stabilizers.

\begin{figure}
\centering
\subfigure{\includegraphics[scale=0.09]{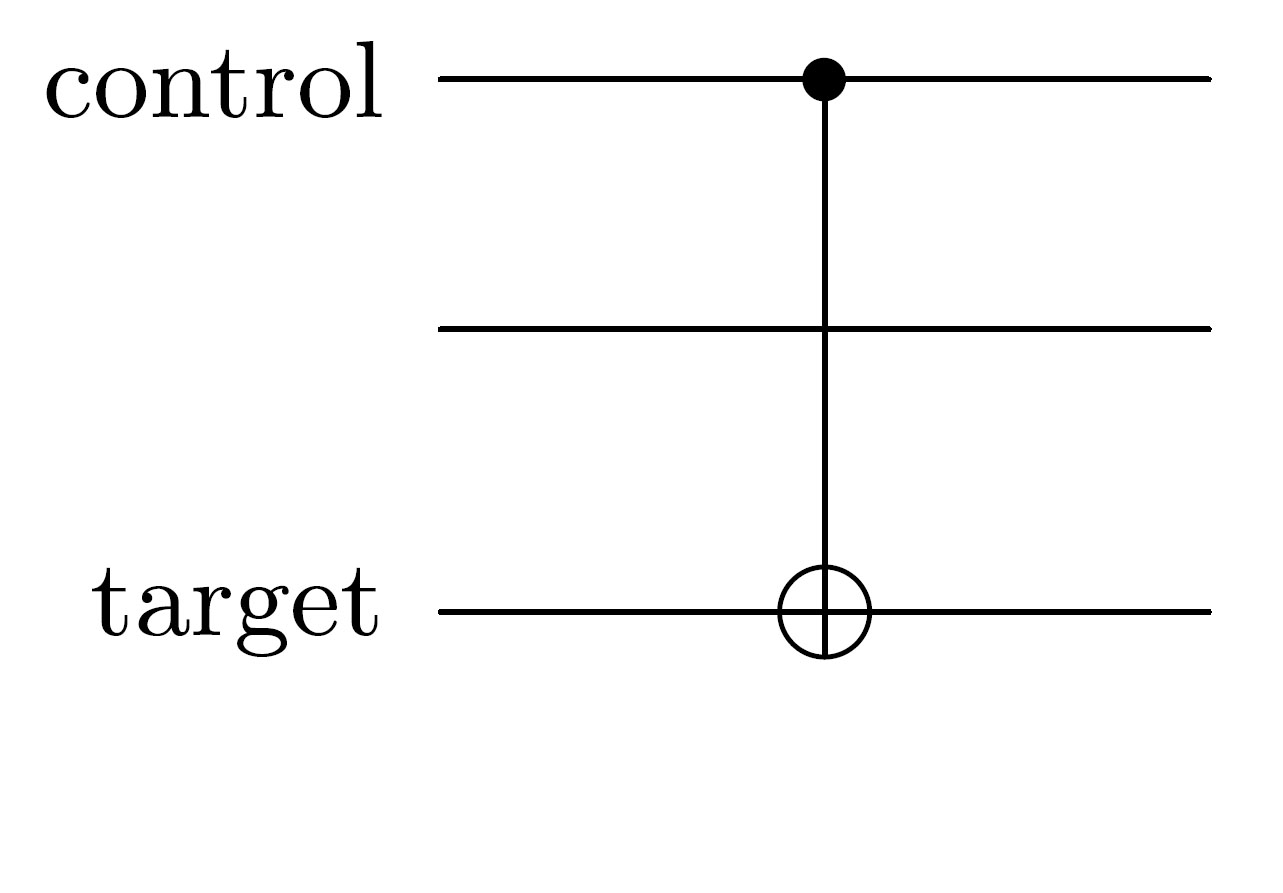}}
\subfigure{\includegraphics[scale=0.08]{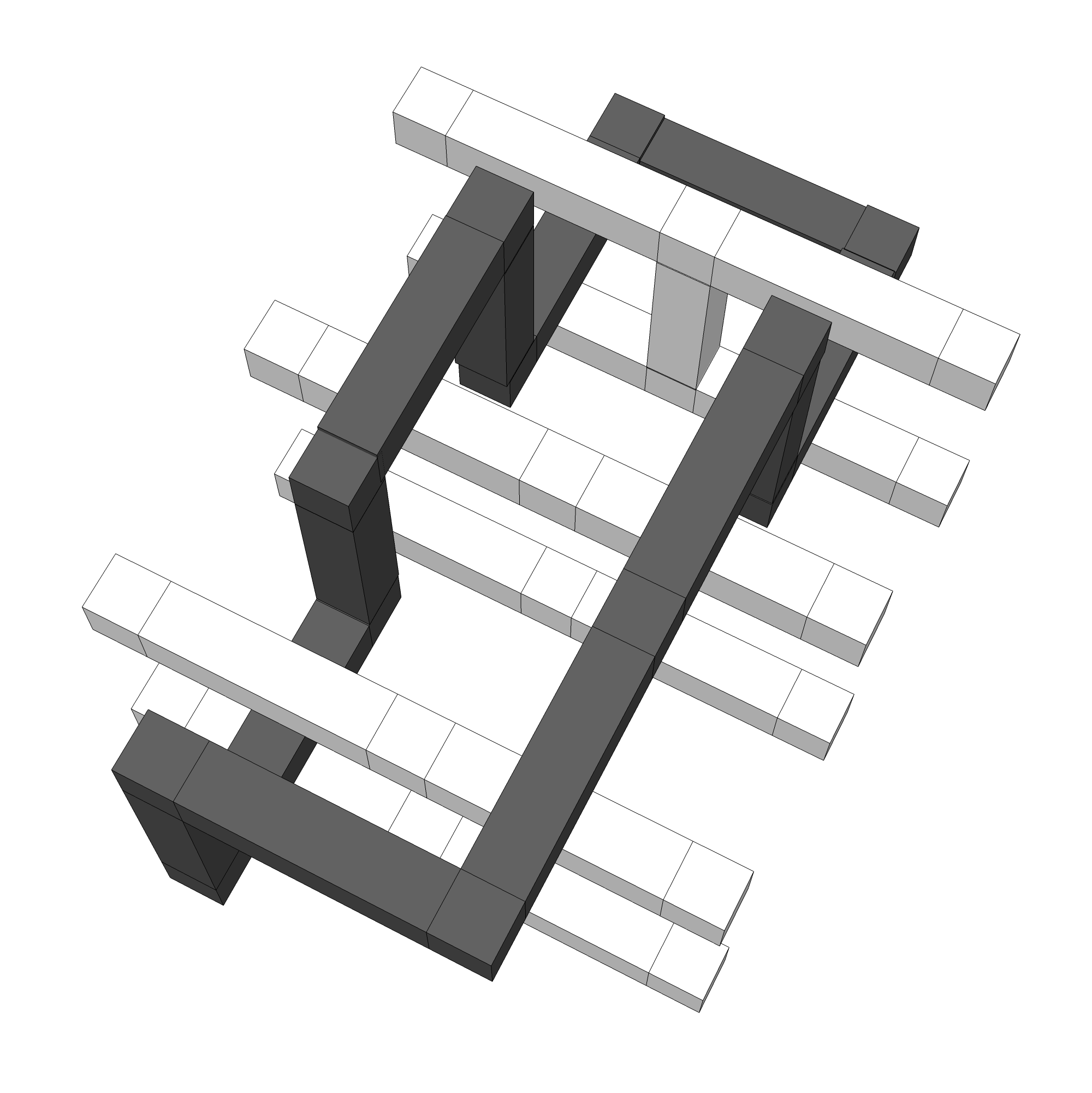}}
\caption{(a) CNOT quantum circuit. (b) Surface code implementation of CNOT using braided defects \cite{Raus07,Raus07d,Fowl12f}. Time runs from left to right. The length of the rectangular structures $d$ is called the code distance and it controls the probability of occurrence of undetected errors, larger $d$ meaning lower error. Each unit of $d$ in a spatial direction represents 2 qubits and in a temporal direction represents a round of surface code error detection.}
\label{scimCN}
\end{figure}

In this work, we limit ourselves to the surface code \cite{Fowl12f}. The surface code represents the simplest topological stabilizer code that can exist, as the number of non-trivial Pauli terms is 4 in any of the stabilizers, the lowest there can be. Fig.~\ref{scimCN} shows a surface code implementation of logical CNOT \cite{Raus07,Raus07d,Fowl12f}. Dark defects are called dual defects and light defects are called primal defects. This structure will perform the same computational process as long as the topology of the structure is preserved. This gives us ways in which we can reduce the volume.

Obtaining the lowest volume structure which is topologically equivalent is important because the volume is a measure of how resource intensive our process is. The probability of undetected errors depends on the length of the individual rectangular structures, called the code distance, which is $d$. Defects also have circumference and separation $d$.

\section{Surface Code Implementation of State Distillation}

A surface code implementation of the circuit in Fig.~\ref{z2circ} is shown in Fig.~\ref{scimz2}. A step by step derivation of this can be found in Appendix A. We define the volume of our structure to be the number of primal cubes contained in the minimum-volume cuboid that can contain the structure. It is straightforward to see that our structure in Fig.~\ref{scimz2} is 7 units long, 2 units wide and 16 units tall and thus has a total volume $V$ = 224. We define a cubical structure of edge length $5d/4$ as a plumbing piece. Thus, we are measuring our volume in units of plumbing pieces. Note that there is a pair of defects for each logical qubit and the 5 layers of CNOT gates has become a single 5 layer dual defect structure.

For the general $|\psi_k\rangle$ distillation case, we will need $2^{k+2}$ qubits, that is $2^{k+3}$ primal defects. The number of layers of plumbing pieces in our family of structures for initialization and CNOTs will be $k+3$, and we need another $k$ layers for $Z_k$ and corrective $Z_{k-1}, Z_{k-2}, ...Z_{1}$ gates. The total volume in plumbing pieces for a single $|\psi_k\rangle$ distillation, assuming all necessary input states are already available, therefore comes out to
\begin{equation}
V_k = 2^{k+3}((k+3)+k)=2^{k+3}(2k+3).
\end{equation}

\begin{figure}
\begin{center}
\includegraphics[width=0.84\linewidth]{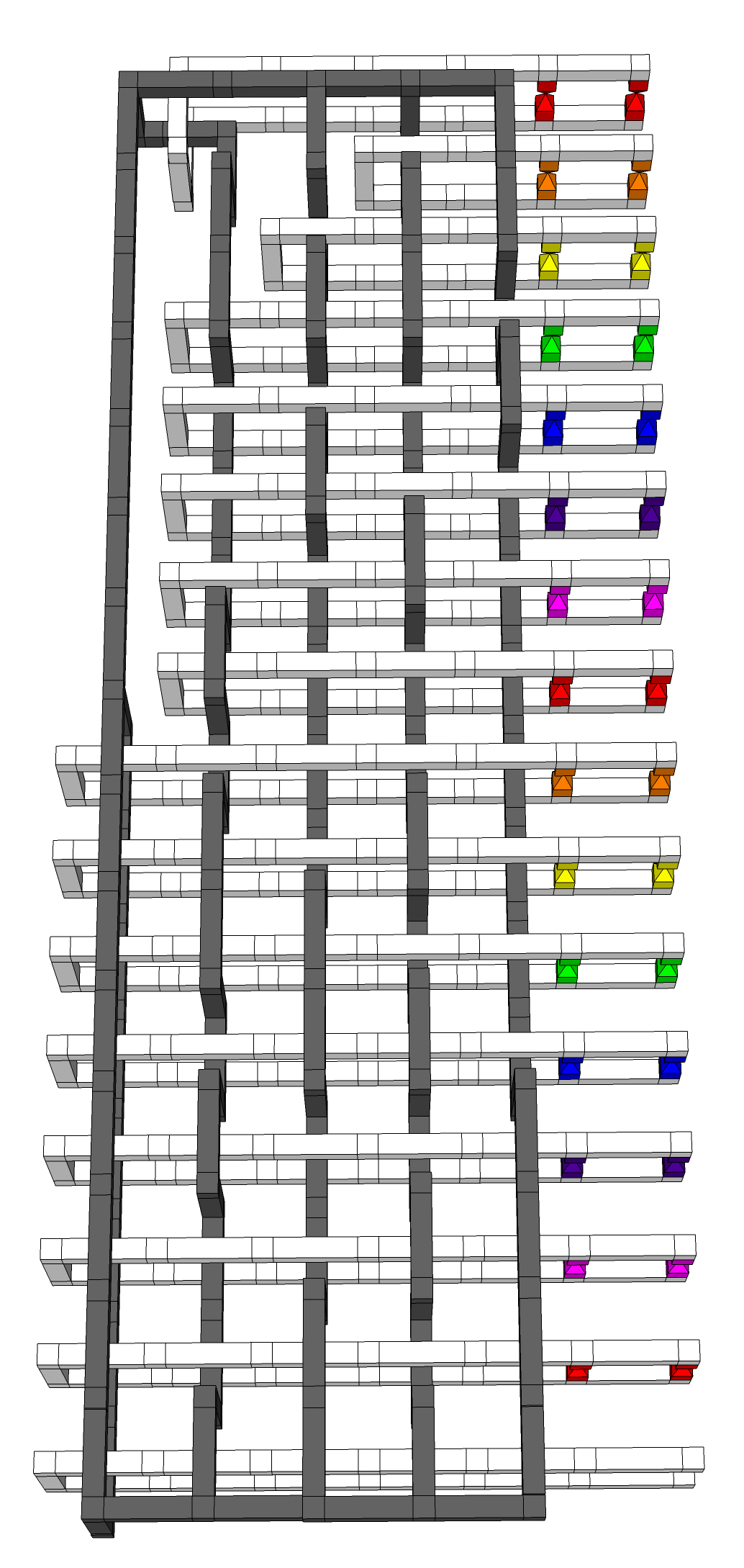}
\end{center}
\caption{Depth 7 cannonical surface-code implementation of Fig.~\ref{z2circ}. A step by step derivation of this can be found in Appendix A.}
\label{scimz2}
\end{figure}

When implementing transversal $Z_k$ gates, we need to leave space for generating all of the $|\psi_j\rangle$, $j\leq k$ states as they add to the total implementation overhead. Note that more highly compressed circuits have been achieved for $k=1$ and $k=2$ \cite{Fowl12h}. However, our primary focus is on $k\ge 3$, and for clear analysis it is useful to have a uniform family of structures for all $k$. Also, the extensive compression techniques used in \cite{Fowl12h} are too laborious to apply to $k\ge 3$.

\section{Overhead Calculation}

\begin{figure}
\begin{center}
\resizebox{85mm}{!}{\includegraphics[viewport=60 55 545 430, clip=true]{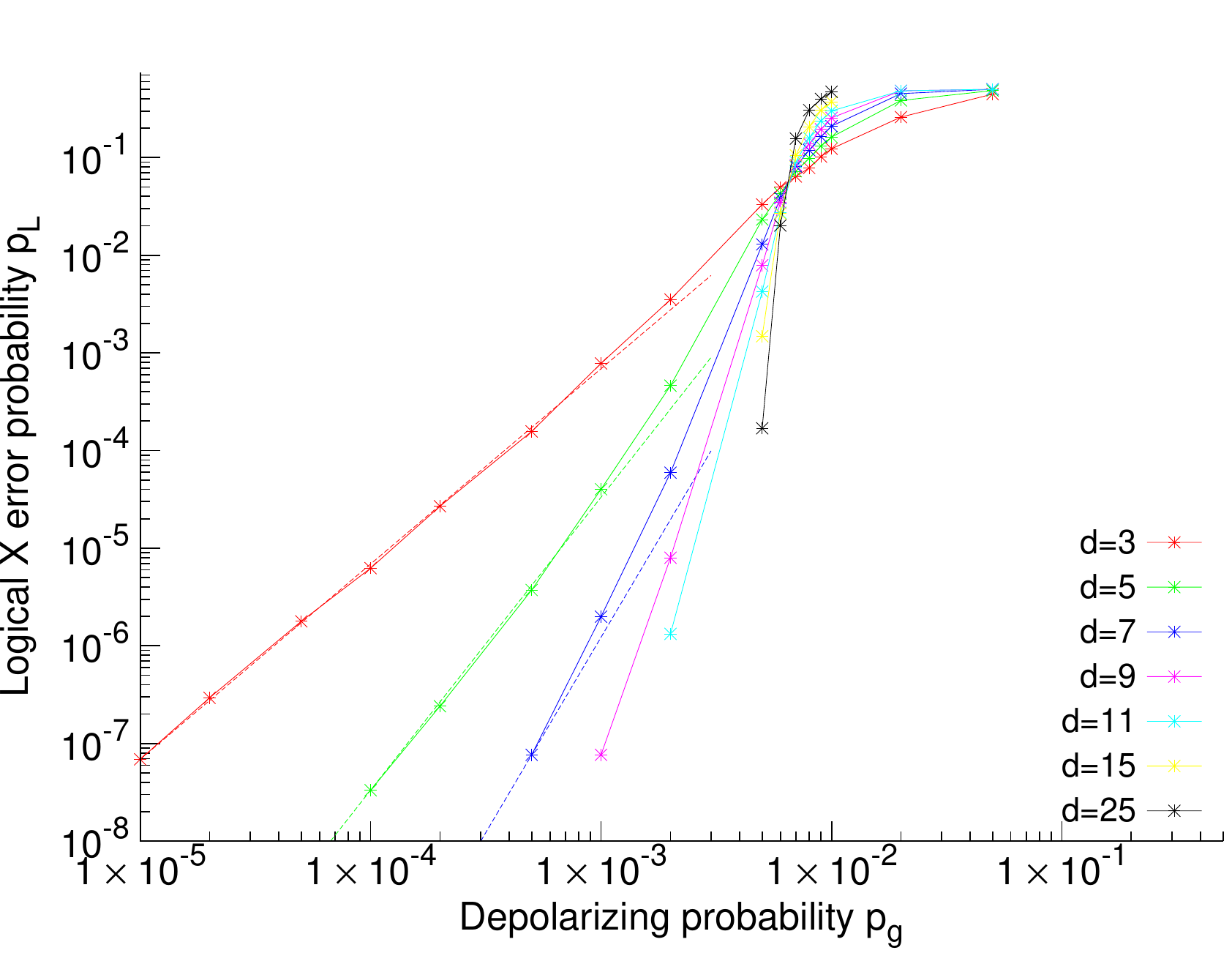}}
\end{center}
\caption{Probability $p_L$ of logical $X$ error per round of surface code error detection for various code distances $d$ and physical gate error rates $p_g$ for correlated fault-tolerant error correction \cite{Fowl13g}. The asymptotic curves are quadratic, cubic, quartic for distances 3, 5, 7 respectively.}
\label{plvsd}
\end{figure}

We now estimate the overhead required to achieve a desired level of fidelity with the distillation protocol described above. The logical error rate of a square patch of surface code of dimension $d\times d$ upper bounds the probability of a logical error connecting neighboring defects and encircling a single defect. From Fig.~\ref{plvsd} \cite{Fowl13g}, we see that the per round probability of logical error of a square surface is $p_L(d,p_g) \sim 0.25 (50p_g)^{(d+1)/2}$. A plumbing piece has $5d/4$ rounds of error detection. There are three distinct classes of logical error that can occur (connecting defects vertically, into the page, or encircling) and, also, a primal and a dual defect can both be contained within a single plumbing piece. From this, the logical error rate of a plumbing piece is upper bounded by
\begin{equation}
P_L(d,p_g) \sim 2\times 3\times\frac{5d}{4}\times p_L(d,p_g) \sim 2d(50p_g)^{\frac{d+1}{2}}.
\end{equation}

If no logical errors are introduced by the distillation circuit, the output error, for a given value of $k$, is $A_kp_s^3$ to lowest order, where $p_s$ is the total error rate of the recursive implementation of $Z_k$ and $A_k$ is given in Table~\ref{valAK}. We assume that the logical circuitry introduces error equal to a fraction $\epsilon$ of the perfect distillation output error, making the actual total output error $p_{out}=(1+\epsilon)A_k p_s^3$. This means that our code distance $d$ must be large enough that the probability of logical error remains under $\epsilon A_kp_s^3$. Given $\epsilon$ and $p_s$, we would look for the lowest value of $d$ which satisfies the condition $V_k P_L(d,p_g) \leq \epsilon A_k p_s^3$. We will actually start with a value of $p_{out}$ and work down, so we choose the lowest value of $d$ such that $V_k P_L(d,p_g) \leq \epsilon p_{out}/(1+\epsilon)$. Note that we allow $\epsilon > 1$.

\begin{table*}
\centering 
\setlength{\extrarowheight}{3pt}
\begin{tabular}{c| c  c  c  c  c  c  c} 
&  &  &  & $k$ &  & \\
\hline 
$p_{out}$  & 1 & 2 & 3 &  4 & 5 & 6 & 7 \\ [0.5ex] 
\hline 
$10^{-5}$ & \ \ \ $2.2 \times 10^{ 5}$ \ \ \ & \ \ \ $7.2 \times 10^{ 5}$ \ \ \ & \ \ \ $3.8 \times 10^{ 6}$ \ \ \ & \ \ \ $1.1 \times 10^{ 7}$ \ \ \ & \ \ \ $2.2 \times 10^{ 9}$ \ \ \ & \ \ \ $1.4 \times 10^{10}$ \ \ \ & \ \ \ $9.8 \times 10^{13}$ \ \ \ \\
$10^{-6}$ & $2.2 \times 10^{5}$ & $1.5 \times 10^{6}$ & $6.6 \times 10^{7}$ & $3.8 \times 10^{8}$ & $2.1 \times 10^{9}$ & $1.3 \times 10^{10}$ & $8.7 \times 10^{13}$ \\
$10^{-7}$ & $4.6 \times 10^{5}$ & $6.1 \times 10^{6}$ & $7.1 \times 10^{7}$ & $3.9 \times 10^{8}$ & $2.1 \times 10^{9}$ & $2.7 \times 10^{10}$ & $1.4 \times 10^{14}$ \\
$10^{-8}$ & $8.4 \times 10^{5}$ & $1.3 \times 10^{7}$ & $7.5 \times 10^{7}$ & $3.9 \times 10^{8}$ & $3.7 \times 10^{9}$ & $3.9 \times 10^{12}$ & $1.4 \times 10^{14}$ \\
$10^{-9}$ & $1.4 \times 10^{6}$ & $1.4 \times 10^{7}$ & $7.5 \times 10^{7}$ & $4.0 \times 10^{8}$ & $4.2 \times 10^{9}$ & $3.8 \times 10^{12}$ & $5.4 \times 10^{16}$ \\
$10^{-10}$ & $1.4 \times 10^{6}$ & $1.6 \times 10^{7}$ & $8.0 \times 10^{7}$ & $7.0 \times 10^{8}$ & $2.9 \times 10^{11}$ & $3.7 \times 10^{12}$ & $5.0 \times 10^{16}$ \\
$10^{-11}$ & $2.9 \times 10^{6}$ & $1.6 \times 10^{7}$ & $8.0 \times 10^{7}$ & $7.7 \times 10^{8}$ & $2.9 \times 10^{11}$ & $3.6 \times 10^{12}$ & $4.6 \times 10^{16}$ \\
$10^{-12}$ & $2.9 \times 10^{6}$ & $1.6 \times 10^{7}$ & $1.4 \times 10^{8}$ & $8.3 \times 10^{8}$ & $2.8 \times 10^{11}$ & $3.6 \times 10^{12}$ & $4.7 \times 10^{16}$ \\
$10^{-13}$ & $3.6 \times 10^{6}$ & $1.9 \times 10^{7}$ & $1.5 \times 10^{8}$ & $2.5 \times 10^{10}$ & $2.8 \times 10^{11}$ & $3.5 \times 10^{12}$ & $4.4 \times 10^{16}$ \\
$10^{-14}$ & $3.6 \times 10^{6}$ & $2.9 \times 10^{7}$ & $1.7 \times 10^{8}$ & $2.5 \times 10^{10}$ & $2.8 \times 10^{11}$ & $6.2 \times 10^{12}$ & $4.3 \times 10^{16}$ \\
$10^{-15}$ & $3.6 \times 10^{6}$ & $3.2 \times 10^{7}$ & $1.8 \times 10^{8}$ & $2.5 \times 10^{10}$ & $2.8 \times 10^{11}$ & $6.1 \times 10^{12}$ & $4.3 \times 10^{16}$ \\
$10^{-16}$ & $4.6 \times 10^{6}$ & $3.4 \times 10^{7}$ & $2.2 \times 10^{9}$ & $2.5 \times 10^{10}$ & $2.8 \times 10^{11}$ & $6.9 \times 10^{12}$ & $7.4 \times 10^{16}$ \\
$10^{-17}$ & $6.3 \times 10^{6}$ & $3.9 \times 10^{7}$ & $2.3 \times 10^{9}$ & $2.6 \times 10^{10}$ & $2.8 \times 10^{11}$ & $1.1 \times 10^{15}$ & $7.1 \times 10^{16}$ \\
$10^{-18}$ & $7.5 \times 10^{6}$ & $3.9 \times 10^{7}$ & $2.4 \times 10^{9}$ & $2.6 \times 10^{10}$ & $2.8 \times 10^{11}$ & $1.0 \times 10^{15}$ & $6.9 \times 10^{16}$ \\
\end{tabular}
\caption{Minimum achieved overheads in qubits-rounds for $p_g=10^{-4}$ and all values of $p_{out}$ of practical interest. Note that, given the approach we have taken and the approximations we have made, $Z_k|+\rangle$ states cannot be obtained at this value of $p_g$ for $k\ge 8$.}
\label{valV4}
\end{table*}

\begin{table}
\centering 
\setlength{\extrarowheight}{3pt}
\begin{tabular}{c| c  c  c  c} 
&  &   \ \ \ \ \ \ \ \ \ \ \ \ \ \ \  \ $k$ &  \\
\hline 
$p_{out}$  & 1 & 2 & 3 &  4  \\ [0.5ex] 
\hline 
$10^{-5}$  & \ $1.5 \times 10^{ 6}$ \ & \ $3.6 \times 10^{ 7}$ \ & \ $4.6 \times 10^{ 8}$ \ & \ \ \ $5.4 \times 10^{11}$ \ \\
$10^{-6}$ & $5.6 \times 10^{6}$ & $5.8 \times 10^{7}$ & $7.6 \times 10^{8}$ & $7.1 \times 10^{11}$ \\
$10^{-7}$ & $5.6 \times 10^{6}$ & $6.4 \times 10^{7}$ & $1.5 \times 10^{10}$ & $5.0 \times 10^{13}$ \\
$10^{-8}$ & $6.7 \times 10^{6}$ & $7.0 \times 10^{7}$ & $1.5 \times 10^{10}$ & $4.4 \times 10^{13}$ \\
$10^{-9}$ & $1.1 \times 10^{7}$ & $1.0 \times 10^{8}$ & $1.4 \times 10^{10}$ & $3.9 \times 10^{13}$ \\
$10^{-10}$ & $1.2 \times 10^{7}$ & $5.7 \times 10^{8}$ & $1.4 \times 10^{10}$ & $3.6 \times 10^{13}$ \\
$10^{-11}$ & $1.4 \times 10^{7}$ & $5.9 \times 10^{8}$ & $1.5 \times 10^{10}$ & $3.3 \times 10^{13}$ \\
$10^{-12}$ & $2.0 \times 10^{7}$ & $5.9 \times 10^{8}$ & $1.5 \times 10^{10}$ & $3.4 \times 10^{13}$ \\
$10^{-13}$ & $2.0 \times 10^{7}$ & $6.3 \times 10^{8}$ & $1.5 \times 10^{10}$ & $3.2 \times 10^{13}$ \\
$10^{-14}$ & $2.3 \times 10^{7}$ & $6.9 \times 10^{8}$ & $1.5 \times 10^{10}$ & $3.0 \times 10^{13}$ \\
$10^{-15}$ & $5.1 \times 10^{7}$ & $9.4 \times 10^{8}$ & $2.2 \times 10^{10}$ & $4.5 \times 10^{13}$ \\
$10^{-16}$ & $5.8 \times 10^{7}$ & $9.5 \times 10^{8}$ & $3.0 \times 10^{11}$ & $4.7 \times 10^{13}$ \\
$10^{-17}$ & $5.8 \times 10^{7}$ & $1.0 \times 10^{9}$ & $3.0 \times 10^{11}$ & $3.3 \times 10^{15}$ \\
$10^{-18}$ & $6.1 \times 10^{7}$ & $1.0 \times 10^{9}$ & $3.3 \times 10^{11}$ & $3.1 \times 10^{15}$ \\
\end{tabular}
\caption{Minimum achieved overheads in qubits-rounds for $p_g=10^{-3}$ and all values of $p_{out}$ of practical interest. Note that, given the approach we have taken and the approximations we have made, $Z_k|+\rangle$ states cannot be obtained at this value of $p_g$ for $k\ge 5$.}
\label{valV3}
\end{table}

In the surface code, we need approximately 10 gates to convert a single qubit state into a logical state \cite{Fowl12f}. We can therefore use $p_{in}=10p_g$ as an approximation for preparing an arbitrary logical state. For $k>1$, we potentially need to perform multiple state injections to achieve the desired rotation before measurement. For a $Z_k$ distillation circuit, on average, we need $Z_{k-1}$ gates half the time, $Z_{k-2}$ gates a quarter of the time, and so on. To account for the probabilistic corrections, we therefore use $p_s= \sum_{i=1}^{k} 2^{-(k-i)} p_{in}$ as a reasonable approximation, where $p_{in}$ is the error rate of implementation of all the required $Z_j$, $j \le k $ rotations. Note that this expression is valid only when the error rates of implementation of the required $Z_j$ rotation gates are all $p_{in}$. In general though, the implementation error rates can be different and we discuss this case later. The relation between output and input error probabilities is therefore given by
\begin{equation}
p_{out} = (1+\epsilon) A_k p_s^3 = (1+\epsilon)A_k (2(1-2^{-k})p_{in})^3.
\end{equation}

The output of state distillation is only kept if no errors are observed. In the limit of low $p_s$ and high $d$, the probability of observing no errors asymptotes to 1. We make the necessary approximation that logical errors due to the failure of error correction are undetectable and always contribute to the output error. Determining the actual fraction of logical errors that are detectable or benign is beyond the capabilities of existing software. We do not expect this approximation to significantly perturb our results. Essentially, this approximation means that occasionally we will be using distillation outputs even though measurements indicate it should be discarded, and our assumed output error rate will be slightly higher than in reality, or equivalently our calculated code distances will be slightly higher than actually required. We also make the reasonable approximation that the error rate of implementing a $Z_k$ gate is the same as that of producing a $|\psi_k\rangle$ state, which is true given sufficient error correction, and equate the volumes, which is a small approximation since the volume of implementing $Z_k$ given $|\psi_k\rangle$ is negligible compared to preparing $|\psi_k\rangle$.

Focusing on errors due to $p_s$, the probability of observing no errors $p_k$ can be lower bounded by the probability of no errors, namely $p_0=(1-p_s)^{n_k}$, where $n_k=2^{k+2}-1$ is the number of qubits. Averaging over many copies of the same distillation circuit, the volume required to ensure successful output is $V/p_0$. For the range of values we will be working with, this lower bound $p_0$ never differs from the exact value $p_k$ by more than 1\%. The exact value is given by the expression \cite{Brav05, Land13}
\begin{equation}
p_k = \frac{1+(2^{k+2}-1)(1-2p_s)^{2^{k+1}}}{2^{k+2}}.
\end{equation}

Let $T_k^{(n)}(p_{out})$ denote the total average volume required to produce a level $n$ distilled $|\psi_k\rangle$ state with error $p_{out}$, including the additional copies required to ensure successful output and all of the lower level and lower $k$ distillation structures. The following recursion relation, which implicitly depends on $\epsilon$ through $p_{in}$, can be used to calculate $T_k^{(n)}(p_{out})$.
\begin{equation}
\begin{cases}
0 & \text{ if } p_{out} \ge 10p_g \\
\frac{1}{p_0}\left(V_k + n_k\sum_{i=1}^{k} \frac{T_i^{(n-1)}(p_{in})}{2^{k-i}}\right) & \text{ if } p_{out} < 10p_g
\end{cases}
\end{equation}

If our required output error rate is greater than $10p_g$, the states can be prepared without distillation and therefore we take the required volume to be zero. In reality, the process of state injection requires significant volume in itself, but equating it to zero here is justified because this volume is still very small compared to the volume needed for the distillation procedure. If the required output error rate is less than $10p_g$, we need to use a $|\psi_k\rangle$ distillation circuit of volume $V_k$, which the first term in the bracket accounts for. Furthermore, as the distillation procedure requires us to implement $n_k$ number of $Z_k$ rotations, we need the same number of $|\psi_k\rangle$ states. Also, to account for the probabilistic corrections while implementing the said $Z_k$ rotations, we need $|\psi_{k-1}\rangle$ states half the time to implement corrective $Z_{k-1}$ rotations, $|\psi_{k-2}\rangle$ states quarter of the time and so on. The volume required to produce these states is what the second factor in the bracket accounts for. The relation between $p_{in}$ and $p_{out}$ is given by equation (6) above.

It should be noted that the input error rates $p_{in}^{(j)}$ of the previous level states $|\psi_j\rangle$ used for implementing $Z_j$ rotations in our circuit need not be equal. If the errors are indeed different, the total error rate of the recursive implementation of $Z_k$ will be $p_s = \sum_{j=1}^{k} p_{in}^{(j)}/2^{k-j}$ and the error rate of our distillation output state $p_{out}$ will be given by $p_{out} = (1+\epsilon) A_k \left( \sum_{j=1}^{k} p_{in}^{(j)}/2^{k-j} \right)^3 $. The removal of this constraint will likely lead to slight further decrease in the total volume required, but we shall not probe this any further. It should be clear from our discussion that given $|\psi_j \rangle$ states for $j\le k$ with error rates $p^{(j)}_{out}$, to implement a $Z_k$ rotation with error $p$ on any arbitrary state $|\phi \rangle$, we would require volume $V =  \sum_{j=1}^{k} T_j(p^{(j)}_{out})/2^{k-j}$, where the numbers $p$ and $p^{(j)}_{out}$ are such that they satisfy $\sum_{j=1}^{k} p^{(j)}_{out}/2^{k-j} \le p$.

Finally, we convert the volume to units of qubits-rounds. As each plumbing piece is $5d/4$ rounds of error detection deep in the temporal direction, and has $5d/4$ data qubits and $5d/4$ measurement qubits in each spatial direction, the ratio of volume in qubits-rounds $V_{qr}$ to volume in plumbing pieces $V_k$ is
\begin{equation}
\eta = \frac{V_{qr}}{V_k} = \frac{5d}{4} \left( \frac{5d}{4} + \frac{5d}{4} \right) \left( \frac{5d}{4}+\frac{5d}{4} \right) = \frac{125}{16} d^3.
\end{equation}

The term $V_k$ in equation (8) has output error probability $p_{out}$, using which we can find the corresponding code distance $d$ as discussed earlier. Replacing the term $V_k$ in equation (8) by $\eta V_k$ gives us the volume in qubits-rounds. For the sake of simplicity, we produce our counts by assuming that all the error probabilities of previous level and lower $k$ circuits are equal, so that equation (6) holds. The general trend we see is that, for a given value of $k$, $p_g$ and $p_{out}$, the overhead decreases with increasing $\epsilon$, suddenly increases at a particular value of $\epsilon$ (indicating an increase in the number of distillation levels), starts decreasing again and so on. We do, however, occasionally find values $\epsilon_1 < \epsilon_2 < \epsilon_3 $ such that overhead$(\epsilon_3) <$ overhead$(\epsilon_1) <$ overhead$(\epsilon_2)$. We calculate the volume in qubits-rounds for a range ($10^{-4}$ to $10^7$) of values of $\epsilon$ and chose the one which results in the lowest overhead. The values have been displayed below in Table~\ref{valV3} and Table~\ref{valV4}.

We must discuss an important point before concluding this section. The fact that we have chosen to keep all our input error probabilities equal decreases the error threshold $p_{th}$. To take an example, the error threshold $p_{th}^{(k)}$ for $|\psi_k \rangle$ distillation circuit is actually greater than $10^{-2}$ for $k \le 5$, but using equation (6) gives us $p_{th}^{(5)}=0.72\times 10^{-2}$. This implies that if $p_g=10^{-3}$, we will not be able to produce a low-error $|\psi_5 \rangle$ state with our assumptions although we certainly can produce that state when the input error probabilities are allowed to be different. Similarly, though it is possible to produce low-error $|\psi_8 \rangle$ and $|\psi_9 \rangle$ states when $p_g=10^{-4}$, with our assumptions we will not be able to. Note, however, that the overhead of producing such states so close to threshold would be unphysically large. The reader should also note that we have taken a top-down approach in our calculations.

\begin{table*}
\centering 
\setlength{\extrarowheight}{3pt}
\begin{tabular}{c| c  c  c  c  c} 
&  &  & $k$ & \\
\hline 
$p_{out}$  & 3 &  4 & 5 & 6 & 7 \\ [0.5ex] 
\hline 
\ \ \ $10^{-5}$ \ \ \ & \ \ \ \  $	2.3\times 10^9$ \ \ \ \  & \ \ \ \  $2.1\times 10^9$ \ \ \ \ & \ \ \ \  $1.9\times 10^9$ \ \ \ \  & \ \ \ \  $2.4\times 10^9$ \ \ \ \ \  & \ \ \ \  $2.2\times 10^9$ \ \ \ \  \\
$10^{-6}$ & $3.0 \times 10^{9}$ & $2.8 \times 10^{9}$ & $2.2 \times 10^{9}$ & $2.7 \times 10^{9}$ & $2.7 \times 10^{9}$ \\
$10^{-7}$ & $4.9 \times 10^{9}$ & $5.0 \times 10^{9}$ & $4.7 \times 10^{9}$ & $4.9 \times 10^{9}$ & $3.2 \times 10^{9}$ \\
$10^{-8}$ & $3.0 \times 10^{10}$ & $3.0 \times 10^{10}$ & $2.9 \times 10^{10}$ & $2.9 \times 10^{10}$ & $3.1 \times 10^{10}$ \\
$10^{-9}$ & $3.2 \times 10^{10}$ & $3.6 \times 10^{10}$ & $3.3 \times 10^{10}$ & $3.0 \times 10^{10}$ & $3.3 \times 10^{10}$ \\
$10^{-10}$ & $3.6 \times 10^{10}$ & $3.8 \times 10^{10}$ & $3.4 \times 10^{10}$ & $3.3 \times 10^{10}$ & $3.4 \times 10^{10}$ \\
$10^{-11}$ & $4.3 \times 10^{10}$ & $4.2 \times 10^{10}$ & $3.8 \times 10^{10}$ & $3.9 \times 10^{10}$ & $4.3 \times 10^{10}$ \\
$10^{-12}$ & $4.9 \times 10^{10}$ & $4.7 \times 10^{10}$ & $5.1 \times 10^{10}$ & $4.7 \times 10^{10}$ & $4.8 \times 10^{10}$ \\
$10^{-13}$ & $7.5 \times 10^{10}$ & $7.6 \times 10^{10}$ & $7.3 \times 10^{10}$ & $7.0 \times 10^{10}$ & $7.4 \times 10^{10}$ \\
$10^{-14}$ & $8.1 \times 10^{10}$ & $7.9 \times 10^{10}$ & $7.9 \times 10^{10}$ & $7.3 \times 10^{10}$ & $7.8 \times 10^{10}$ \\
$10^{-15}$ & $8.8 \times 10^{10}$ & $8.6 \times 10^{10}$ & $8.8 \times 10^{10}$ & $8.5 \times 10^{10}$ & $9.1 \times 10^{10}$ \\
$10^{-16}$ & $9.2 \times 10^{10}$ & $9.5 \times 10^{10}$ & $9.1 \times 10^{10}$ & $9.7 \times 10^{10}$ & $9.5 \times 10^{10}$ \\
$10^{-17}$ & $9.9 \times 10^{10}$ & $1.1 \times 10^{11}$ & $1.1 \times 10^{11}$ & $1.1 \times 10^{11}$ & $1.1 \times 10^{11}$ \\
$10^{-18}$ & $1.1 \times 10^{11}$ & $1.1 \times 10^{11}$ & $1.1 \times 10^{11}$ & $1.1 \times 10^{11}$ & $1.1 \times 10^{11}$ \\
\end{tabular}
\caption{Minimum achieved overheads in qubits-rounds using approximating sequences for all values of $p_{out}$ of practical interest, and $p_g=10^{-3}$. Note that the volume remains almost the same for any particular $p_{out}$ irrespective of $k$.} 
\label{valV5}
\end{table*}

\begin{table*}
\centering 
\setlength{\extrarowheight}{3pt}
\begin{tabular}{c| c  c  c  c  c} 
&  &  & $k$ & \\
\hline 
$p_{out}$  & 3 &  4 & 5 & 6 & 7 \\ [0.5ex] 
\hline 
\ \ \ $10^{-5}$ \ \ \ & \ \ \ \  $	5.8\times 10^7$ \ \ \ \  & \ \ \ \  $5.2\times 10^7$ \ \ \ \ & \ \ \ \  $4.8\times 10^7$ \ \ \ \  & \ \ \ \  $6.1\times 10^7$ \ \ \ \ \  & \ \ \ \  $5.5\times 10^7$ \ \ \ \  \\
$10^{-6}$ & $5.2 \times 10^{8}$ & $5.0 \times 10^{8}$ & $3.8 \times 10^{8}$ & $4.7 \times 10^{8}$ & $4.7 \times 10^{8}$ \\
$10^{-7}$ & $6.6 \times 10^{8}$ & $6.8 \times 10^{8}$ & $6.3 \times 10^{8}$ & $6.6 \times 10^{8}$ & $5.9 \times 10^{8}$ \\
$10^{-8}$ & $7.6 \times 10^{8}$ & $7.5 \times 10^{8}$ & $7.3 \times 10^{8}$ & $7.3 \times 10^{8}$ & $7.8 \times 10^{8}$ \\
$10^{-9}$ & $8.9 \times 10^{8}$ & $9.9 \times 10^{8}$ & $9.0 \times 10^{8}$ & $8.2 \times 10^{8}$ & $9.0 \times 10^{8}$ \\
$10^{-10}$ & $9.9 \times 10^{8}$ & $1.0 \times 10^{9}$ & $9.2 \times 10^{8}$ & $8.9 \times 10^{8}$ & $9.4 \times 10^{8}$ \\
$10^{-11}$ & $1.3 \times 10^{9}$ & $1.2 \times 10^{9}$ & $11.1 \times 10^{8}$ & $1.2 \times 10^{9}$ & $1.3 \times 10^{9}$ \\
$10^{-12}$ & $1.4 \times 10^{9}$ & $1.3 \times 10^{9}$ & $1.4 \times 10^{9}$ & $1.3 \times 10^{9}$ & $1.3 \times 10^{9}$ \\
$10^{-13}$ & $2.5 \times 10^{9}$ & $2.6 \times 10^{9}$ & $2.5 \times 10^{9}$ & $2.4 \times 10^{9}$ & $2.5 \times 10^{9}$ \\
$10^{-14}$ & $2.9 \times 10^{9}$ & $2.8 \times 10^{9}$ & $2.8 \times 10^{9}$ & $2.6 \times 10^{9}$ & $2.8 \times 10^{9}$ \\
$10^{-15}$ & $3.4 \times 10^{9}$ & $3.3 \times 10^{9}$ & $3.4 \times 10^{9}$ & $3.3 \times 10^{9}$ & $3.5 \times 10^{9}$ \\
$10^{-16}$ & $3.5 \times 10^{9}$ & $3.6 \times 10^{9}$ & $3.5 \times 10^{9}$ & $3.7 \times 10^{9}$ & $3.6 \times 10^{9}$ \\
$10^{-17}$ & $1.0 \times 10^{10}$ & $1.1 \times 10^{10}$ & $1.1 \times 10^{10}$ & $1.1 \times 10^{10}$ & $1.1 \times 10^{10}$ \\
$10^{-18}$ & $1.1 \times 10^{10}$ & $1.2 \times 10^{10}$ & $1.2 \times 10^{10}$ & $1.2 \times 10^{10}$ & $1.2 \times 10^{10}$ \\
\end{tabular}
\caption{Minimum achieved overheads in qubits-rounds using approximating sequences for all values of $p_{out}$ of practical interest, and $p_g=10^{-4}$. Note that the volume remains almost the same for any particular $p_{out}$ irrespective of $k$.} 
\label{valV6}
\end{table*}

We end this section by working out an example which illustrates how we have calculated the numbers. Let us take $p_{in}=10^{-3}$, $p_{out}=10^{-8}$ and $k=2$. This means that the $|\psi_1\rangle$ and $|\psi_2\rangle$ states required for implementing the transversal $Z_2$ gates must each have error probability below $10^{-8}/1.5$. We find that the value of $\epsilon$ which gives us the net minimum volume is $1.41$. We first calculate the volume it takes to prepare the $|\psi_2\rangle$ state. For the given values, we find that the code distance $d$ such that $V_k P_L(d,p_g) \leq \epsilon p_{out}/(1+\epsilon)$ i.e. the lowest $d$ which satisfies $224 P_L(d,10^{-3}) \leq 1.94 \times 10^{-9}$ is $d=19$. Also, $p_{in}=\sqrt[3]{p_{out}/A_k(\epsilon+1)}/(2(1-2^{-k}))=2.86\times 10^{-4}$ and $1/p_0=1.0065$. As $p_{in}<10p_g$, we need additional lower distillation levels to achieve the required output error. Continuing similarly, for the lower level $k=2$ circuit, $p_{in}=10^{-2}$, $d'=11$ and $1/p'_0=1.255$; for the corrective $k=1$ circuit, $p_{in}=2.57\times 10^{-2}$, $d''=11$ and $1/p''_0=1.12$. As both these input errors are greater than $10p_g$, we need no more distillation. Using $\eta=125d^3/16$, the volume $V$ required to produce the $|\psi_2 \rangle$ state is equal to
\[ \frac{1}{p_0} \left(\eta V_2 + 15 \left( \frac{\eta' V_2}{p'_0} + \frac{(0.5)\eta'' V_1}{p''_0} \right) \right) = 6.37\times10^7\].
For preparing the $|\psi_1\rangle$ state, we need two distillation levels, with $d=19$, $p_{in}=7.34\times10^{-4}$ and $1/p_0=1.005$ for the top level; $d=11$, $p_{in}=3.52\times10^{-2}$ and $1/p_0=1.285$ for the bottom level. The volume for this is $1.18\times 10^7$ qubits-rounds and the total volume, therefore, comes out to $6.96\times 10^7$ qubits-rounds.

\section{Approximating Sequences}

Quantum algorithms are implemented by successively applying quantum gates, which are unitary operators acting on the state the gate is being applied to. For instance, a single qubit unitary in $SU(2)$ is a $2 \times 2$ matrix of the form

\begin{equation}
U = \begin{pmatrix}
e^{i\alpha} \cos(\theta) & e^{i\beta} \sin(\theta) \\
-e^{-i\beta} \sin(\theta) & e^{-i\alpha} \cos(\theta)
\end{pmatrix}.
\end{equation}

Distinguishing between this continuous spectrum of unitaries is impossible given the presence of noise. Also, quantum error correction requires us to only work with a discrete number of gates. It turns out that while we cannot execute all quantum gates perfectly, we can approximate them to an arbitrary accuracy with a finite set of fundamental gates. Clifford gates plus the $T$ gate make one such set. The Solovay-Kitaev theorem \cite{Daws04b} states that any single qubit unitary can be approximated to precision $\delta$ by using $O(\log^c (\frac{1}{\delta}))$ gates from this set.

As the overhead requirement of a $T$ gate is considerably more than other gates, looking for approximate sequences with low $T$ gate counts is of great practical interest. We concern ourselves with a recent work by Kliuchnikov, Maslov and Mosca \cite{Kliu12} where an algorithm has been proposed which uses an optimal number of $T$ gates. Their algorithm uses $O(\log (\frac{1}{\delta}))$ number of $T$ gates to approximate a unitary to precision $\delta$.

The work states that any set of unitary matrices obtained by sequentially applying gates from the Clifford group and $T$ gate is equivalent to the set of unitary matrices with entries in the ring $\mathbb{Z}[\frac{1}{\sqrt{2}},i]$, which is the set of numbers $\{\sum(a_n+b_ni)(\frac{1}{\sqrt{2}})^n|a,b \in \mathbb{Z}\}$ where the sum is finite. It further shows that, in case where an exact decomposition is possible, repeatedly applying carefully chosen sequence of gates from the set $\{H,HT,HT^2,HT^3\}$ always leads us to a small set of unitaries whose decomposition is already known. So if we want to decompose a unitary $U$, we know that we can always choose $k_1, k_2...k_n$ such that $HT^{k_1}HT^{k_2}...HT^{k_n}U=U'$, where the decomposition of $U'$ is known. We thus have $U=HT^{-k_n}...HT^{-k_2}HT^{-k_1}U'$. Other approximation methods exist \cite{Ross14}.

In this work, we are concerned with approximating $Z_k$ rotations, as defined earlier, which may not necessarily have an exact decomposition. This case is dealt with in another work \cite{Kliu12a} where, to approximate the $Z_k$ matrix, entries are chosen from the ring $\mathbb{Z}[\omega]$ where $\omega=e^{i\pi /8}$, which is a subset of $\mathbb{Z}[\frac{1}{\sqrt{2}},i]$. They are chosen so as to minimize the approximation error and to satisfy unitarity, and the conditions that the row and column should be unit vectors. Once such a matrix is obtained, it is decomposed into a sequence containing Clifford group and $T$ gate as previously described.

Suppose the unitary $U$ is approximated by a sequence $U_a$. We measure the precision with which $U_a$ approximates $U$ by the metric \cite{Fowl04c}
\begin{equation}
\delta (U,U_a) = \sqrt{\frac{2-|\text{tr}(U^\dagger U_a) |}{2}}
\end{equation}

We find that if $U$ and $U_a$ are sufficiently close so that $\delta (U, U_a)$ is very small, error probability is
\begin{equation}
p=1-|\langle +|U^\dagger U_a|+ \rangle|^2 \approx 2(\delta (U, U_a))^2
\end{equation}

We estimate the overhead by making the assumption that $T$ gates are the major contributor and therefore, we can ignore the overhead requirements of other gates without much effect on our resource count. We further assume that the error is only introduced in our circuit by $T$ gates and the rest of the gates are error free. Therefore, given a number $n$ of $T$ gates with error $p$ each, to approximate a gate $U_a$ which further approximates a gate $U$, the net error in our approximation will be $np+2(\delta (U, U_a))^2$. It is straightforward to see that if the overhead of a $T$ gate with error $p$ is $V(p)$, then the overhead estimate in the given case will be $nV(p)$.

For a given total error, we calculate the overhead for various approximate sequences, each with a different number of $T$ gates and value of $\delta (U,U_a)$, and select the one which results in the lowest net overhead. The calculated lowest overheads have been displayed in Table~\ref{valV5} and Table~\ref{valV6}. The approximate sequences have been obtained from the SQCT software page (http://code.google.com/p/sqct/) maintained by authors of the work we have described.

\section{Discussion}

We have calculated the overhead required in implementing a particular set of $Z$ rotations, first implemented through direct distillation \cite{Land13}, then by using generated approximate sequences \cite{Kliu12,Kliu12a}. The general trend that we see is that for a given value of $p_g$, there is a $k_0$ such that approximate sequences almost always lead to orders of magnitude lower overhead for $k>k_0$. For $p_g=10^{-3}$, $k_0=3$ and for $p_g=10^{-4}$, $k_0=4$. This trend is observed because the overhead keeps on increasing with $k$ in case of direct distillation; whereas, given a number $n$ of $T$ gates, the family of sequences generated to approximate $Z_k$ rotations lead to more or less the same precision irrespective of $k$. Therefore, if we use approximate sequences, the overhead will depend only on the output precision and will be independent of $k$. For $p_g=10^{-4}$, $p_{out}=10^{-18}$ and $k=7$, direct distillation was found to take up to 6 orders of magnitude more overhead. One must note that the value of $k$ above which approximate sequences become more efficient depends on $p_g$.

\section{Acknowledgements}
\label{ack}

AGF acknowledges funding from the US Office of the Director of National Intelligence (ODNI), Intelligence Advanced Research Projects Activity (IARPA), through the US Army Research Office grant No. W911NF-10-1-0334, and partial support from the Australian Research Council Centre of Excellence for Quantum Computation and Communication Technology (CE110001027) and the U.S. Army Research Office (W911NF-13-1-0024). All statements of fact, opinion or conclusions contained herein are those of the authors and should not be construed as representing the official views or policies of IARPA, the ODNI, or the US Government.

\bibliography{References}

\appendix

\begin{figure*}
\section{Step-by-step compression of $Z_2|+\rangle$ distillation circuit}
\vspace{3mm}
\begin{center}
\includegraphics[width=\linewidth]{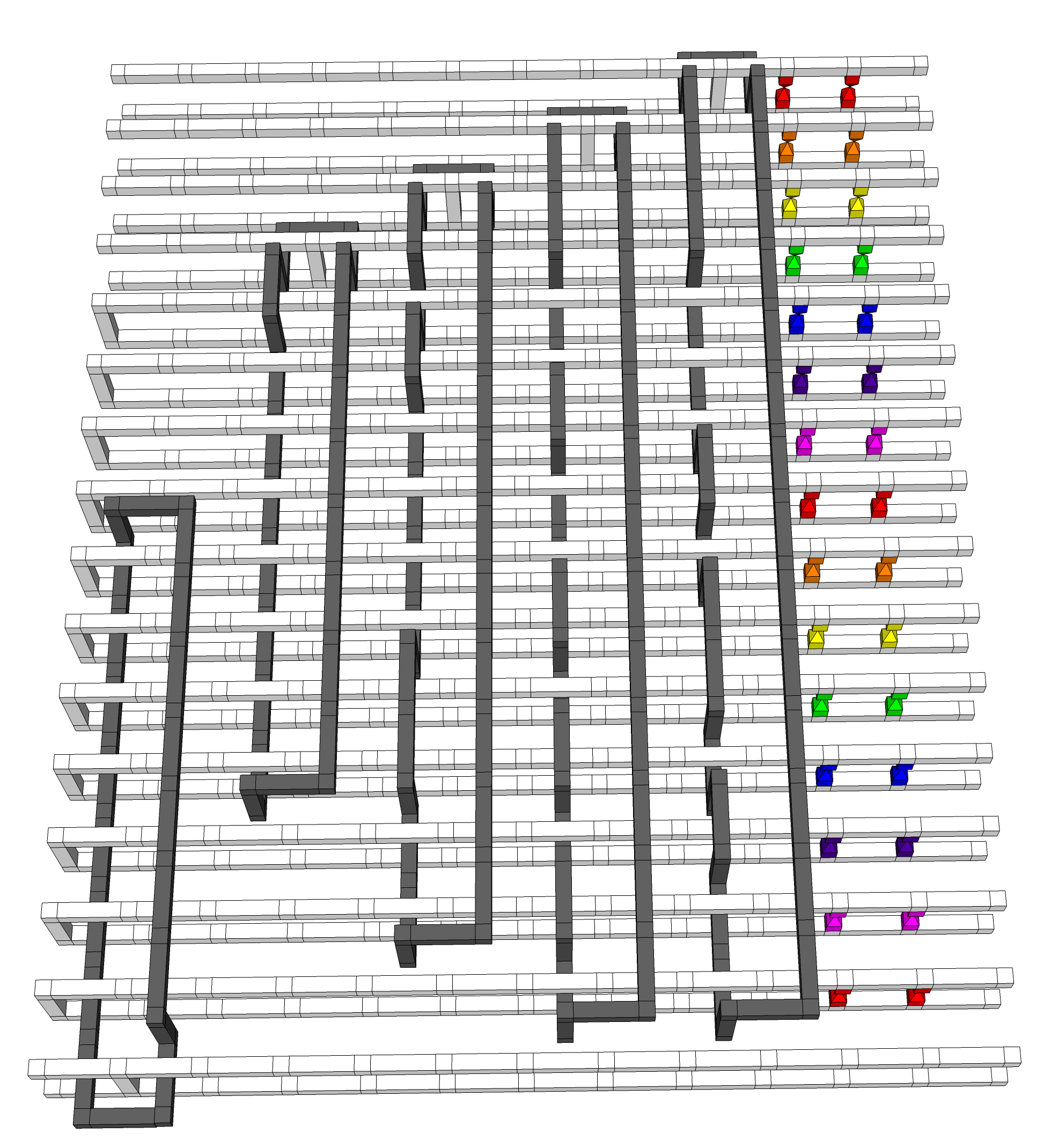}
\end{center}
\caption{Depth 13 canonical surface code implementation of Fig.~\ref{z2circ}. Dark structures are called dual defects, light structures are called primal defects. The depth is defined to be the maximum number of small primal cubes from left to right.}
\label{app1}
\end{figure*}

\begin{figure*}
\begin{center}
\includegraphics[width=\linewidth]{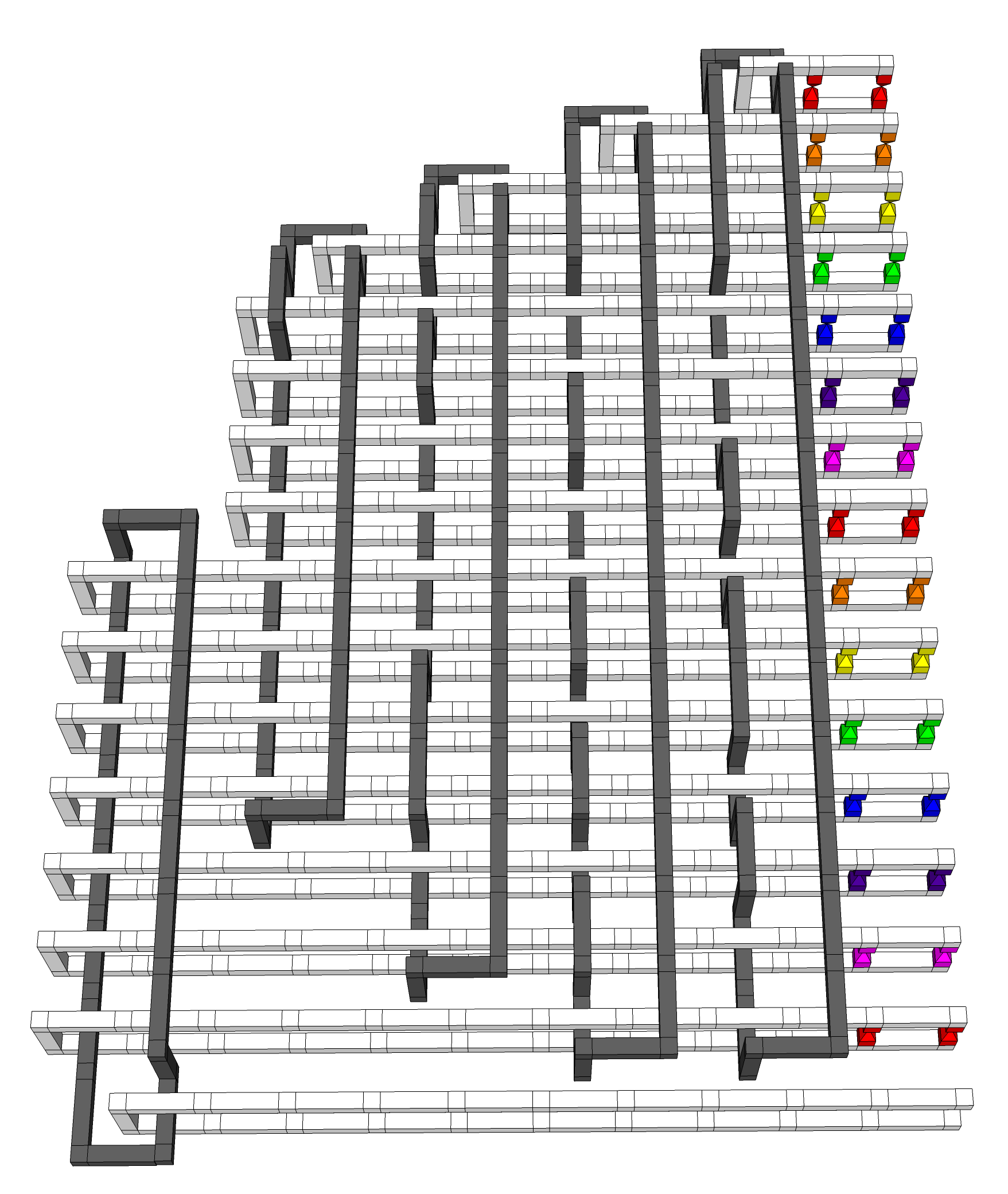}
\end{center}
\caption{Initialization patterns and bumps representing X basis measurement have been pushed in as far as possible.}
\label{app2}
\end{figure*}

\begin{figure*}
\begin{center}
\includegraphics[width=0.95\linewidth]{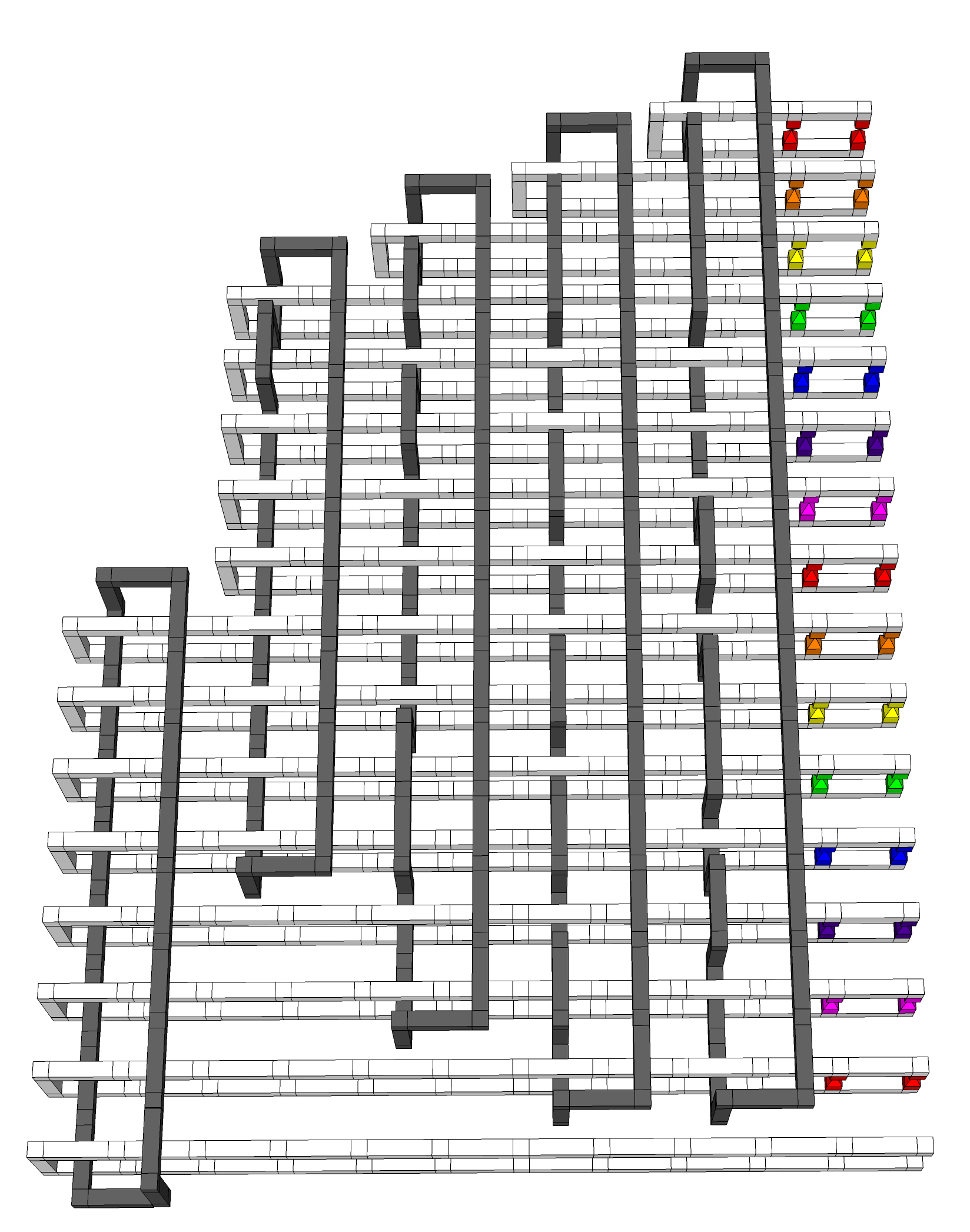}
\end{center}
\caption{Control ends of all dual defects have been reversed.}
\label{app3}
\end{figure*}

\begin{figure*}
\begin{center}
\includegraphics[width=0.95\linewidth]{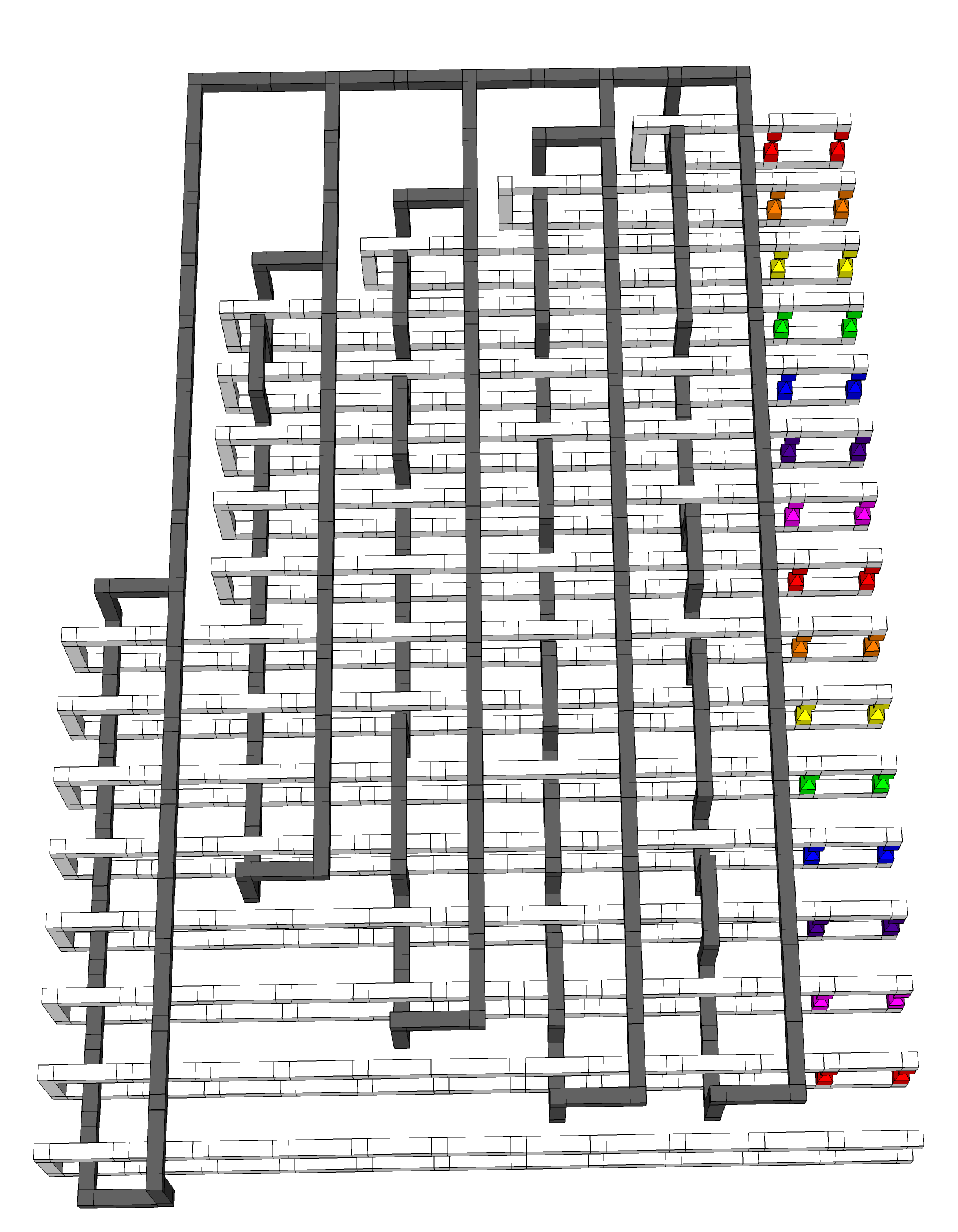}
\end{center}
\caption{Four bridges \cite{Fowl12h} have been simultaneously inserted at the top between the dual defects. This can be shown to implement the same computation.}\label{app4}
\end{figure*}

\begin{figure*}
\begin{center}
\includegraphics[width=0.55\linewidth]{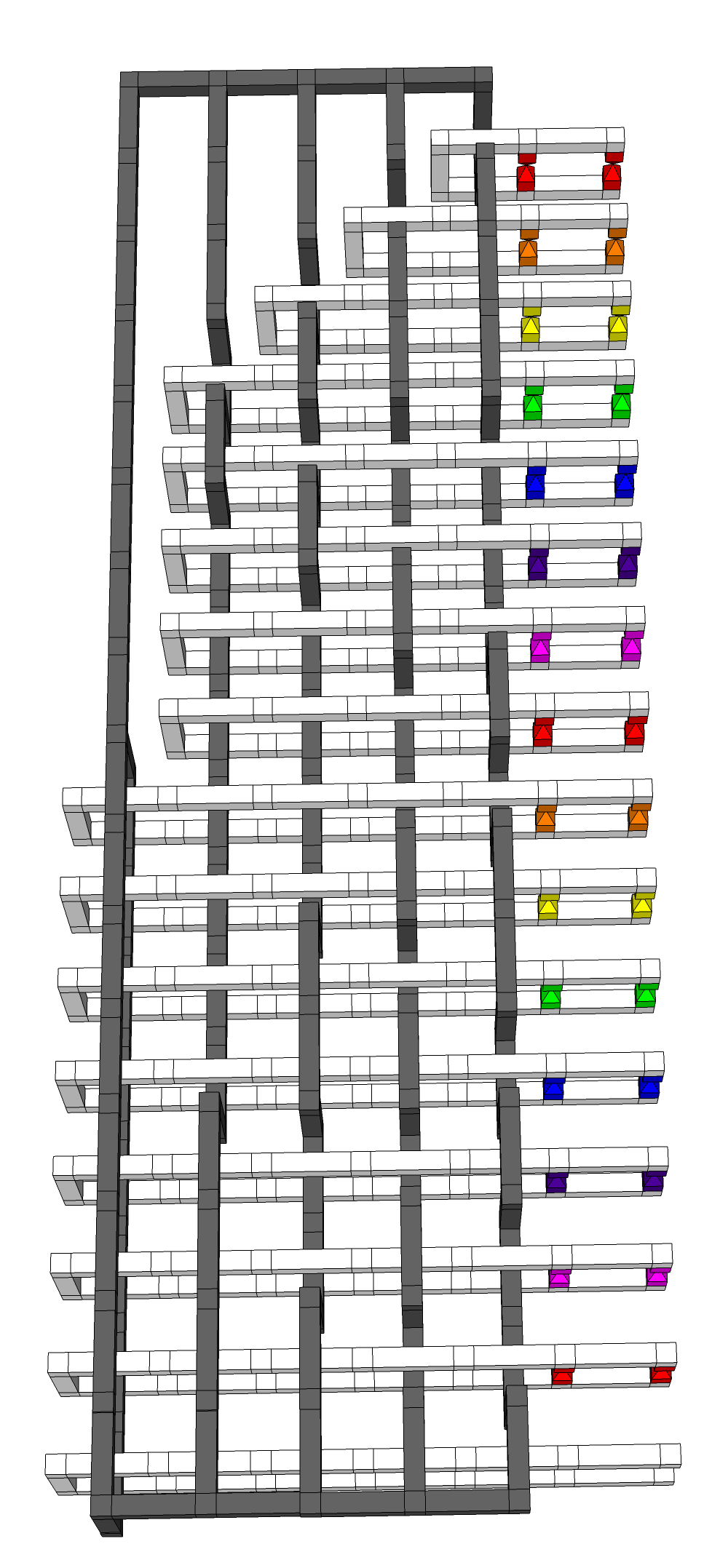}
\end{center}
\caption{The right halves of the four rightmost dual defects have been dragged across the bridges to the bottom, allowing the whole structure to be compressed to depth 7.}
\label{app5}
\end{figure*}

\begin{figure*}
\begin{center}
\includegraphics[width=0.55\linewidth]{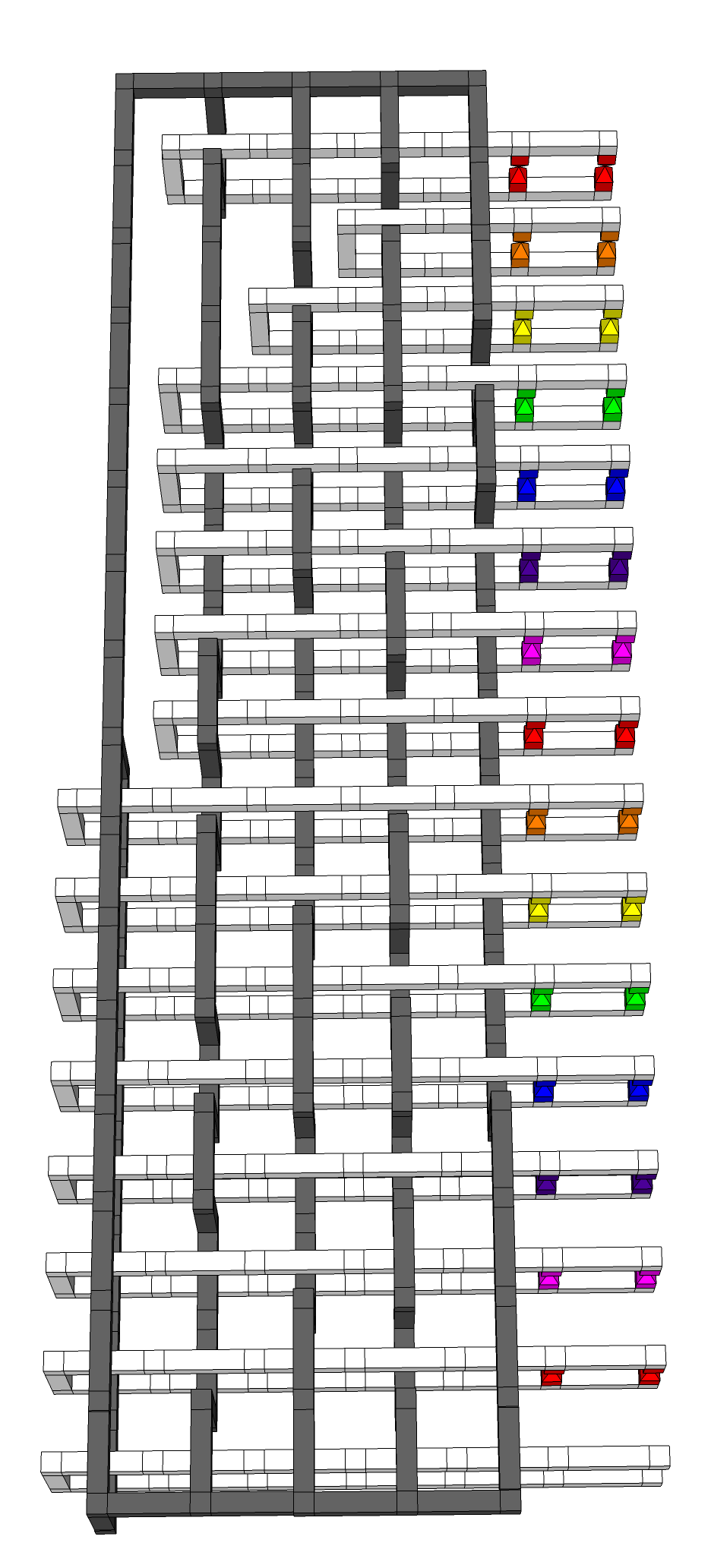}
\end{center}
\caption{Second and last dual defects have been exchanged to facilitate further compression.}
\label{app6}
\end{figure*}

\begin{figure*}
\begin{center}
\includegraphics[width=0.58\linewidth]{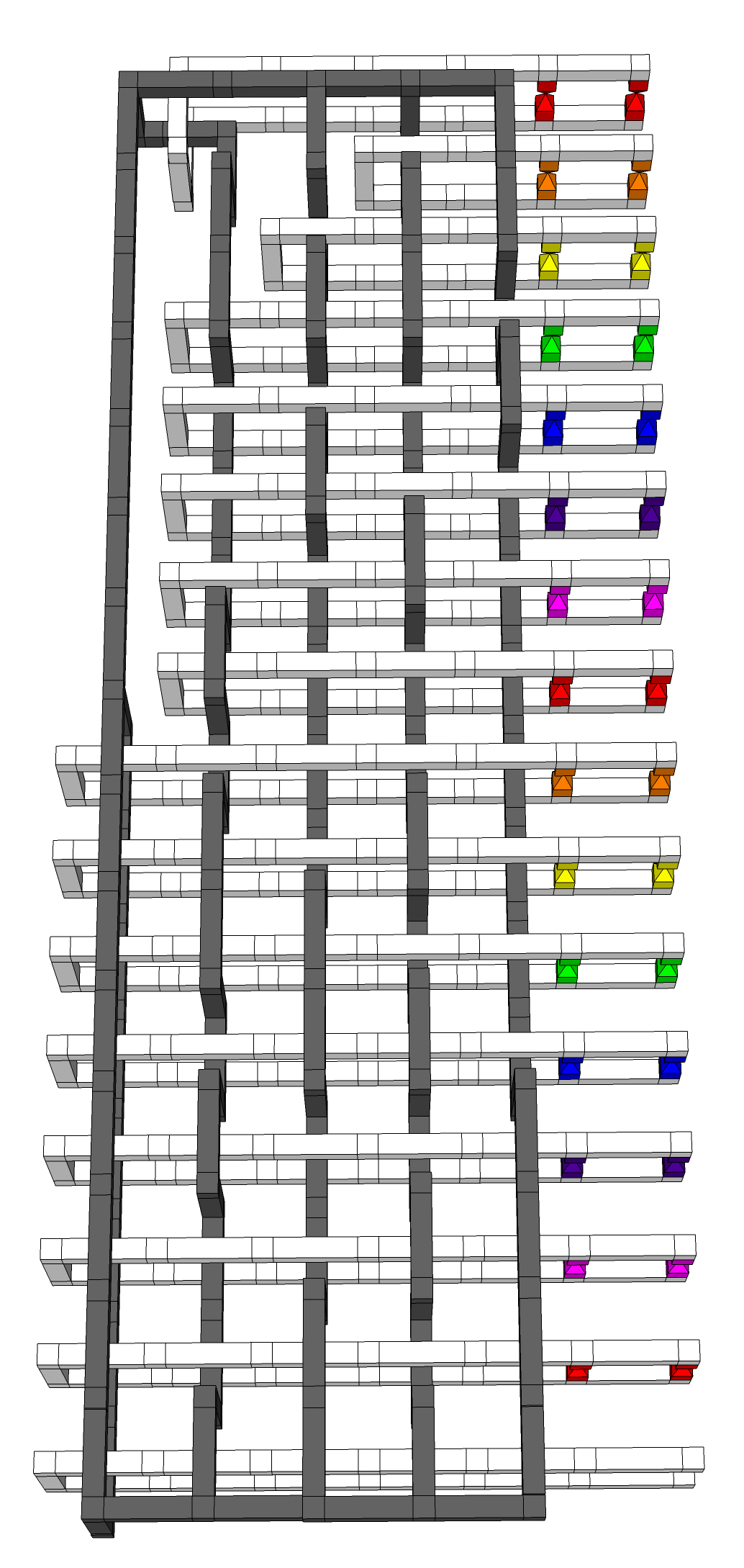}
\end{center}
\caption{Minor topological deformations have been made at the top to reduce the height by one unit. We have now arrived at Fig.~\ref{scimz2}}
\label{app7}
\end{figure*}

\end{document}